\documentclass[superscriptaddress,aps,epsfig,nofootinbib]{revtex4}
\usepackage{graphicx}
\usepackage{epstopdf}
\usepackage{latexsym}
\usepackage{amssymb}
\usepackage{textcomp}
\usepackage{commath}
\usepackage{amsmath}
\usepackage[center]{subfigure}
\usepackage{feynmp}
\usepackage{array}
\makeatletter
\DeclareRobustCommand{\cev}[1]{%
  \mathpalette\do@cev{#1}%
}
\newcommand{\do@cev}[2]{%
  \fix@cev{#1}{+}%
  \reflectbox{$\m@th#1\vec{\reflectbox{$\fix@cev{#1}{-}\m@th#1#2\fix@cev{#1}{+}$}}$}%
  \fix@cev{#1}{-}%
}
\newcommand{\fix@cev}[2]{%
  \ifx#1\displaystyle
    \mkern#23mu
  \else
    \ifx#1\textstyle
      \mkern#23mu
    \else
      \ifx#1\scriptstyle
        \mkern#22mu
      \else
        \mkern#22mu
      \fi
    \fi
  \fi
}

\makeatother

\begin{document}
\renewcommand\thesection{\arabic{section}}
\renewcommand\thesubsection{\thesection.\arabic{subsection}}
\renewcommand\thesubsubsection{\thesubsection.\arabic{subsubsection}}
 \newcommand{\bq}{\begin{equation}}
 \newcommand{\eq}{\end{equation}}
 \newcommand{\bqn}{\begin{eqnarray}}
 \newcommand{\eqn}{\end{eqnarray}}
 \newcommand{\nb}{\nonumber}
 \newcommand{\lb}{\label}
 \newcommand{\glgt}{g_{\text{LGT}}}
 \newcommand{\Y}{{\cal{Y}}}
 \newcommand{\B}{{\cal{B}}} 
 \newcommand{\A}{{\cal{A}}} 
 
 \title{Path Integral Quantization of Lorentz Gauge Theory of Gravity:\\
 With a Proof of Unitarity and Full Renormalizability in the Vacuum}
 \author{Ahmad Borzou}
\email{ahmad_borzou@baylor.edu}

\affiliation{EUCOS-CASPER, Physics Department, Baylor University, Waco, TX 76798-7316, USA}

\date{\today}

\begin{abstract}
We show that a spinless theory of gravity is also allowed by the kinematics of general relativity. 
In the absence of fermions the spinless theory of gravity and the theories in the standard model of particle physics are the same Yang-Mills theories with different gauge groups. Therefore, every theorem of a pure Yang-Mills theory is valid for the spinless theory of gravity in vacuum, i.e. it is unitary and renormalizable to all orders. 
When fermions are present the spinless theory of gravity has an extra constraint due to the tetrad postulate. A path integral quantization of this theory and all the Feynman rules are presented.
\end{abstract}

\maketitle

\section{Introduction}
Gravity is expected to be fundamentally quantized \cite{Eppley1977,DeWitt2011,Albers2008,Kiefer2014}. 
A successful quantization of any theory depends on whether it is at the same time unitary and renormalizable to all orders of perturbation. General relativity is the widely accepted theory of gravity at this time. It has been shown that even in the absence of matter it is not renormalizable. Additional terms can be added to make it renormalizable but at the cost of violating unitarity \cite{tHooft1974,Deser1999}. In this paper we show that Lorentz gauge theory of gravity (LGT) is a spinless pure Yang-Mills theory of gravity that at least in the absence of matter is both unitary and renormalizable to all orders of perturbation.

Unification has been the origin of several breakthroughs in physics. Newton's laws of motion,  Maxwell's equations, the two theories of relativity, and the standard electroweak theory, are all in favor of the idea that the laws of physics are everywhere the same and the observed differences are just details. 
A conclusion that can be drawn is that whenever different patterns are observed in two theories in physics, one pattern is either wrong or an asymptote of the other.

While the theories in the standard model of particle physics are Yang-Mills, general relativity is not. Although the two are gauge theories, there are fundamental differences.
If Einstein had not finished his general theory of relativity but the standard model was at its current formulation, 
how would we develop a theory of gravity that also respects the general covariance and the equivalence principles?
This question has been asked by Feynman \cite{Feynman2003} and later also by others \cite{Weinberg1965,Deser1975}. The result has been that the most simple theory that one can build for a spin 2 and massless particle will be general relativity.
Nevertheless, spin 2 and zero mass are the initial assumptions and experiment is the only way to justify an assumption in physics. 
Although there is only an experimental upper limit on the mass of gravitons, in LGT also the mass is assumed zero.
Moreover, current observations can not uniquely determine the spin of gravitons.  
Gravitons can not be spin 0 because otherwise a coupling between gravity and light will not be possible. 
Gravitons can not be spin 1 or we should have seen repulsive forces of gravity in the same length scale as we see the attractive ones.
On the other hand, theories with spin higher than 2 are still being investigated \cite{Vasiliev1992,Konstein2000, Ammon2011, Castro2016}. 
In this paper we show that if a unifying scheme is followed, i.e. if gravitons are to be represented by the gauge field of a Yang-Mills theory---in the same way the gauge bosons are represented by the gauge fields in the standard model---spinless gravitons will be our choice.

A massless Yang-Mills theory is uniquely defined if its gauge group is specified. The equivalence principle suggests that the gauge group for gravity should have a universal nature, i.e. either the Poincare or the Lorentz groups are our best options. 
Interestingly, both of the symmetries are locally preserved in the kinematics of general relativity. For this reason the same kinematics is assumed in LGT. 
A thorough description of the kinematics can be found in \cite{WeinbergGRBook}. 
The action of the matter is invariant under arbitrary coordinate transformations in the space-time and also under arbitrary Lorentz transformations of the spinors in the tangent spaces. 
The two symmetries are independent and while the former is associated with conservation of energy-momentum tensor, the latter introduces a conserved Lorentz current to be derived in section \ref{Sec:symmetry}. 
In the same section we show that if the energy-momentum tensor is the source of gravity, the metric will be dynamical and therefore the spin of gravitons will be 2. On the other hand if the Lorentz current is the source of gravity, the dynamical variable will be the spin connections and gravitons will be spinless.
In the former case the metric as the dynamical variable is a tensor and not a gauge field of coordinate transformations and therefore general relativity is not Yang-Mills. In the latter case the spin connection as the dynamical variable is not a tensor but a gauge field of the internal Lorentz transformations and consequently the resulting theory---LGT---will be Yang-Mills.

In the end we would like to mention that some of the open problems in high energy physics are the direct consequences of the assumption that gravitons are spin 2 and will no longer exist if gravitons are spinless. 
First, the vacuum energy must gravitate in a theory with spin 2 graviton. There is however no observation in favor of such gravitational field \cite{Weinberg1989,Martin2012}. Vacuum energy will not go into the source of a spinless theory and does not gravitate. 
Second, the space-time is fundamentally quantized in a spin 2 theory of gravity. But, a quantized time is not consistent with quantum mechanics \cite{Kiefer2013,Isham1993}. In a spinless theory of gravity time is just a background.
Third, the coupling constant in a spin 2 theory has negative engineering dimension which means that even the first condition for renormalizability would not be passed \cite{Weinberg1980}. In a spinless theory however the coupling constant is dimensionless.  

The structure of this article is as follows. In section \ref{Sec:symmetry} the dynamics for a spinless and massless theory of gravity is set. Section \ref{Sec_Review} is devoted to a brief review of LGT and some of its applications. In section \ref{Sec:Quantization_NoMatter} it is shown that LGT in the absence of fermions falls exactly under the category of general Yang-Mills theories and every theorem that is proved to be valid in those theories is also valid in LGT. Notably, the theory is unitary and renormalizable to all orders. 
In section \ref{Sec:Quantization_WithMatter} it is shown that in the presence of matter LGT has an extra constraint in comparison with the Yang-Mills theories of the standard model. LGT is subsequently quantized under the extra constraint using the path integral method. The effective Lagrangian is derived and the Feynman rules are presented. In section \ref{Sec:conclusion} a conclusion is given.

\section{The Dynamics of a Massless Spinless Theory of Gravity}
\lb{Sec:symmetry}
It is well known that conserved currents are uniquely determined by symmetries and if a conserved current is assumed to be the source in a theory, the dynamics of the theory is also uniquely specified by that assumption. In the following we first review the subject in known theories and then derive the dynamics of LGT as a massless spinless theory of gravity. 

We start with a SU(N) gauge theory. The system is defined by the following action
\bqn
\lb{Eq:action_general}
S = \int e d^4x \left(\frac{i}{2}\bar{\psi}\gamma^{l}e_l^{~\mu}D_{\mu}\psi -\frac{i}{2}\bar{\psi}\cev{D}_{\mu}\gamma^{l}e_l^{~\mu}\psi\right),
\eqn
where $e_{i\mu} \equiv \hat{e}_i \cdot \hat{e}_{\mu}$ is the dot product between the tetrad and unit vectors tangent to space-time coordinates.
Here and in the following the Latin and Greek indices belong to the internal Lorentz space and space-time respectively.
Only the spinor term and the gauge field change under an infinitesimal SU(N) transformation while the tetrad remains the same
$\delta\psi = g t^a\alpha^a\psi$, $\delta A_{~\mu}^{a} = D_{\mu}\alpha^a$, and $\delta e_{i\mu} = 0$
where $t^a$ are the generators of the SU(N), and $\alpha^a$ are arbitrary parameters. 
Invariance of the action under a SU(N) transformation induces the conserved current. 
The change in the action will be $\delta S = \frac{\partial S}{\partial \psi}\delta \psi + \frac{\partial S}{\partial A^a_{\mu}}\delta A^a_{\mu}$.
The firs term however will be zero by the spinor field equations. Therefore, 
$\delta S = \int e d^4x \frac{\partial {\cal{L}}_{M}}{\partial A^a_{\mu}} D_{\mu}\alpha^a$.
After neglecting the surface terms it reads
$\delta S = - \int e d^4x D_{\mu}\frac{\partial {\cal{L}}_{M}}{\partial A^a_{\mu}} \alpha^a=0$.
So,
$D_{\mu}\frac{\partial {\cal{L}}_{M}}{\partial A^a_{\mu}} = 0$,
and the conserved current is simply
\bqn
J^{a\mu} = \frac{\partial {\cal{L}}_{M}}{\partial A^a_{\mu}}.
\eqn
If this current is the source in a theory, the matter Lagrangian should be varied with respect to $A^a_{\mu}$ in order to derive the same current. Therefore, $A^a_{\mu}$ is the dynamical variable of the theory.  

Although the previous example looked trivial, it does not in theories based on space-time symmetries. 
The action is again given by equation \eqref{Eq:action_general}---with different definition for covariant derivatives---and is invariant under general coordinate transformation and Lorentz transformations in the tangent spaces \cite{WeinbergGRBook}. 
The dual local symmetries means that any object can be dual classified. For example, the metric is a second-rank tensor under coordinate transformations but a scalar under internal Lorentz transformations while the spin connection is a vector under the former and a gaug field, not a tensor, under the latter \cite{WeinbergGRBook}. 
A tetrad $e_{i\mu}$---more accurately it is the dot product of the tetrad with the coordinate frame---is a vector in both spaces and can be decomposed using the equivalence principle that guarantees a unique---up to a global transformation---free falling frame, i.e. a unique set of four orthogonal unit vectors, at each point of the space-time. It can be easily shown that the tetrad can be written as
\bqn
\lb{tetrad_decomposed}
e_{i\mu} = \eta^{\bar{m}\bar{n}}e_{\bar{m}i}e_{\bar{n}\mu},
\eqn
where a bar refers to the unique free falling frame, $e_{\bar{n}\mu}$ are the components of the unique frame in an arbitrary coordinate frame, and $e_{\bar{m}i}$ are the components of the unique frame in an arbitrary internal Lorentz frame.

To derive the conservation of energy-momentum tensor and also the dynamics of a theory in which energy-momentum tensor is the source, we study the invariance of the action under $x^{\mu'} = x^{\mu} + \zeta^{\mu}$
\bqn
\int e'd^4x' {\cal{L}}_{M}'\left(x'\right) = \int e d^4x \left(\frac{\delta {\cal{L}}_{M}}{\delta e_{m\alpha}}\delta e_{m\alpha} + \partial_{\mu}{\cal{L}}_{M}\zeta^{\mu}\right),
\eqn
where the volume element is invariant itself, a variation with respect to the spinor field is dropped by its field equation, the spin connection $A_{ij\mu}$ is written in terms of the tetrad using the tetrad postulate, and the variation in the tetrad is in a single location
$
\delta e_{m\alpha} \equiv e_{m\alpha}'(x) - e_{m\alpha}(x) = e_{m\alpha}'(x') - e_{m\alpha}(x) -\partial_{\mu}e_{m\alpha}\zeta^{\mu}=-e_{m\mu}\partial_{\alpha}\zeta^{\mu}-\partial_{\mu}e_{m\alpha}\zeta^{\mu}.
$
After defining $t^{m\alpha} \equiv \frac{\delta {\cal{L}}_{M}}{\delta e_{m\alpha}}$, a substitution reads
$
\int e d^4x \left( \frac{1}{e}\partial_{\alpha}\left(et^{m\alpha}e_{m\mu}\right)-\partial_{\mu}e_{m\alpha}t^{m\alpha}+\partial_{\mu}{\cal{L}}_{M}\right)\zeta^{\mu}=0.
$
For a symmetric $t_{\alpha\lambda}$ and using the fact that $\partial_{\mu}\delta^{\lambda}_{\alpha}=\partial_{\mu}\left(e^{i\lambda}e_{i\alpha}\right)=0$ and $\partial_{\mu}g^{\alpha\lambda}=-\Gamma^{\alpha}_{\mu\beta}g^{\beta\lambda}-\Gamma^{\lambda}_{\mu\beta}g^{\alpha\beta}$ 
and also using the arbitrariness of $\zeta^{\mu}$ one can conclude the conservation of the energy-momentum tensor 
\bqn
D_{\alpha} T_{\mu}^{\alpha} \equiv D_{\alpha}\left(t_{\mu}^{\alpha}+g_{\mu}^{\alpha}{\cal{L}}_{M}\right)=0,
\eqn
where $D_{\alpha}$ is interchangeable with the pseudo-Riemannian covariant derivative $\nabla_{\alpha}$ when it is acting on a scalar of the internal Lorentz frame. 
It is now easy to show that if the energy-momentum tensor is the source in a theory, as in general relativity, the dynamical variable will be $e_{\bar{n}\mu}$ in equation \eqref{tetrad_decomposed}. 
On the other hand, if the action is varied with respect to $e_{\bar{m}i}$, the energy-momentum tensor will not be obtained
because the metric can be written as $g_{\mu\nu} = \eta^{\bar{m}\bar{n}}e_{\bar{m}\mu}e_{\bar{n}\nu}$ and is not a function of $e_{\bar{m}i}$---it also shows that the metric is a scalar of the Lorentz space---and therefore
\bqn
\lb{Var_wrt_e_im}
\frac{\delta g_{\mu\nu}}{\delta e_{\bar{m}i}} =0,
~~\frac{\delta \Gamma^{\alpha}_{\mu\beta}}{\delta e_{\bar{m}i}} =0,
~~\frac{\delta \sqrt{-g}}{\delta e_{\bar{m}i}} = 0,
~~\frac{\delta e}{\delta e_{\bar{m}i}} =0.
\eqn
Consequently, ${\cal{L}}_{M}\delta e/\delta e_{\bar{m}i}$ will no longer lead to the second term in the energy-momentum tensor.

Finally, we would like to derive the conserved Lorentz current through the invariance under internal Lorentz transformations
\bqn
\delta S = \int e d^4x \frac{\delta {\cal{L}}_{M}}{\delta A_{ij\mu}} \delta A_{ij\mu} = 0,
\eqn
where a variation in the spinor field is dropped again by means of the field equations, and the tetrad is written in terms of the spin connection using the tetrad postulate. Under an internal Lorentz transformation the spin connection changes as
\bqn
\lb{Eq:LorVarOfA}
\delta A_{ij\mu} = D_{\mu}\omega_{ij}, 
\eqn
and since $\omega_{ij}$ is arbitrary, the Lorentz current $J^{ij\mu}=\delta {\cal{L}}_{M}/\delta A_{ij\mu}$ is conserved $D_{\mu}J^{ij\mu}=0$. 
If we assume that this current is the source of a theory, one can easily show that $e_{\bar{m}\mu}$ in equation \eqref{tetrad_decomposed} is not dynamical. This can be realized by a deeper look at the tetrad postulate
\bqn
\partial_{\mu}e_{i\nu}-\Gamma_{\mu\nu}^{\rho}e_{i\rho}-A_{ij\mu}e^j_{~\nu}=0.
\eqn
A general variation in the spin connection therefore can be written as 
\bqn
\delta A_{ij\mu} = \delta_1 A_{ij\mu} + \delta_2 A_{ij\mu} = 
D_{\mu}\left(e_j^{~\nu}\delta_1 e_{i\nu}\right)-\delta_1\Gamma_{\mu\nu}^{\rho}e_{i\rho} + D_{\mu}\left(e_j^{~\nu}\delta_2 e_{i\nu}\right),
\eqn
where $\delta_1 \equiv \delta/\delta e_{\bar{m}\mu}$ and $\delta_2 \equiv \delta/\delta e_{\bar{m}i}$, and equation \eqref{Var_wrt_e_im} is used.
Comparing with equation \eqref{Eq:LorVarOfA}, it is clear that if the Lorentz current is assumed to be the source in a theory, as in LGT, one has to vary the action with respect to $e_{\bar{m}i}$.
On the other, because of $\delta_1\Gamma_{\mu\nu}^{\rho}$ a variation with respect to $e_{\bar{m}\mu}$ leads to something different than the conserved Lorentz current---in contradiction with our initial assumption.  
Consequently, equation \eqref{Var_wrt_e_im} implies that the metric and its determinant and the Christoffel symbols are all background fields under the dynamics of LGT.
This in the path integral language means that the three mentioned variables remain the same on every path that the dynamical variable of the theory $A_{ij\mu}$ takes. 
This feature is exactly the same as in the Yang-Mills theories of the standard model of particle physics and as will be shown in the following sections will enable us to show that every theorem that is proven for general Yang-Mills theories is also valid in the vacuum version of LGT. 
Since the dynamical variable is the gauge field and not a tensor of the internal Lorentz group, it is spinless. However, its existence is the reason that spin of other particles is locally preserved.

\section{A Review of Lorentz Gauge Theory}
\lb{Sec_Review}
The construction starts with the action for matter in equation \eqref{Eq:action_general}. 
It is invariant under two independent transformations. The changes under general covariance are
\bqn
\psi'(x')=\psi(x), ~ ~e_{i\mu}'(x')=\frac{\partial x^{\nu}}{\partial x^{\mu'}}e_{i\nu}(x),~ ~A_{ij\mu}'(x')=\frac{\partial x^{\nu}}{\partial x^{\mu'}}A_{ij\nu}(x),
\eqn
while the changes under internal Lorentz transformations are
\bqn
\lb{Eq:LorTransform}
&&\psi'(x) = \Lambda_{\frac{1}{2}}\psi(x), ~  ~
e_{i\mu}'(x) = \Lambda_i^{~j} e_{j\mu}(x),~ ~
A_{ij\mu}'(x) = \Lambda_i^{~m}\Lambda_j^{~n}A_{mn\mu}(x) + \partial_{\mu}\Lambda_i^{~m}\Lambda_j^{~n}\eta_{mn},
\eqn
where
\bqn
&&\Lambda_{\frac{1}{2}} = e^{\frac{g}{2} S^{mn}\omega_{mn}},\nb\\
&&\Lambda_i^{~j} = \left(e^{\frac{g}{2} J^{mn}\omega_{mn}}\right)_i^{~j},
\eqn
and $S^{mn}$ are the generators of the Lorentz group in the spinor representation and in terms of the Dirac matrices can be written as $S^{mn}=\frac{1}{4}[\gamma^m,\gamma^n]$ and $\left(J^{mn}\right)_{ij}=\delta^m_i\delta^n_j-\delta^m_j\delta^n_i$ are the generators of the Lorentz group in the vector representation. The Lie algebra reads
\bqn
\lb{Eq:LorLieAlg}
[{\cal{J}}^{mn},{\cal{J}}^{ij}] = \eta^{ni}{\cal{J}}^{mj} + \eta^{mj}{\cal{J}}^{ni} - \eta^{nj}{\cal{J}}^{mi} - \eta^{mi}{\cal{J}}^{nj},
\eqn
with ${\cal{J}}$ is the generator of the Lorentz group in any representation. The covariant derivative is defined to locally preserve both of the symmetries
\bqn
\lb{Eq:LorCovDer}
&&D_{\mu}\psi = \partial_{\mu}\psi - \frac{g}{2}A_{mn\mu}S^{mn}\psi,\\
&&D_{\mu}A_{\alpha_1\cdots\alpha_k}^{\beta_1\cdots\beta_l}=\nabla_{\mu}A_{\alpha_1\cdots\alpha_k}^{\beta_1\cdots\beta_l},
\eqn
with $\nabla$ being the covariant derivative in a pseudo-Riemannian space-time. 
The covariant derivative in the tensor representation of the Lorentz group can be found
by applying the derivative on an arbitrary Lorentz tensor like $\bar{\psi}\gamma_m\psi$, the product rule, the derivative in the spinor representation, and finally the following
\bqn
[\gamma^i,S^{mn}] = \eta^{im}\gamma^n - \eta^{in}\gamma^m.
\eqn
An important example would be
\bqn
\lb{Eq:CovDerVec}
D_{\mu}\omega_{mn} = \partial_{\mu}\omega_{mn} - gA_{m~\mu}^{~k}\omega_{kn} - gA_{n~\mu}^{~k}\omega_{mk}.
\eqn
Another important concept in this kinematics is the tetrad postulate
\bqn
D_{\mu}e_{i\nu} = \partial_{\mu}e_{i\nu} - \Gamma^{\rho}_{\mu\nu}e_{i\rho}-gA_{i~\mu}^{~j}e_{j\nu} = 0,
\eqn
which can be derived by the fact that the partial derivative of the tetrad in a free falling frame is zero. In this frame the coordinate and the Lorentz unit vectors coincide. An arbitrary coordinate and an arbitrary internal Lorentz transformation from the free falling frame results in the equation above. 

So far the kinematics of general relativity---with fermions included---is reviewed. More detailes can be found in \cite{WeinbergGRBook}. 
The same kinematics is kept for LGT. The difference between the two theories is the assumption regarding the spin of the gravitons or equivalently the source of gravity. 
As was discussed in section \ref{Sec:symmetry}, our kinematics allows two assumptions of spin 2 or spinless and the latter is the choice for LGT. 
Under this assumption the gravitons will be represented by gauge fields rather than tensors and the theory is Yang-Mills. Therefore, from the unification perspective a spinless theory of gravity is better than a theory based on spin 2 assumption.

Lorentz gauge theory is defined by \cite{Borzou2016}
\bqn
\lb{action}
S&=&\int e d^4x\Big[{\cal{L}}_{\text{LGT}}+{\cal{L}}_{M}+{\cal{L}}_{C}\Big].
\eqn
Here the Lagrangian for matter is given in equation \eqref{Eq:action_general} and the Lagrangian for the Lorentz guage field is given by 
\bqn
\lb{NLA}
{\cal{L}}_{\text{LGT}}&=&-\frac{1}{4}F_{\mu \nu ij}F^{\mu\nu ij},
\eqn
where $F_{\mu \nu ij}$ is the strength tensor of the internal Lorentz space and can be converted to the Riemann curvature tensor through
\bqn
R_{\mu\nu \rho \sigma} = e^i_{~\rho} e^j_{~\sigma} F_{\mu\nu ij}.
\eqn
It is important to note that the Lagrangian above is a 2nd derivative rather than a higher derivative term since the dynamical variable is the spin connection rather than the metric. 

Since the tetrad and the spin connection are dependent through the tetrad postulate, either the tetrad should be expressed in terms of the gauge field or the constraint should be inserted to the action by means of the Lagrange multiplier method.
This is the last term in the action of LGT and is given by 
\bqn
{\cal{L}}_{\text{C}} = S^{\mu\nu m}D_{\mu}e_{m\nu},
\eqn
in which the tetrad postulate $D_{\mu}e_{m\nu}=0$ is coupled to the multiplier.

Variation of the action with respect to the Lagrange multiplier leads to the tetrad postulate again.
A variation with respect to the gauge field yields
\bqn
\lb{fielddynamic}
&&D_{\nu}F^{\mu \nu mn}=J^{m n \mu},\nb\\
&&J^{m n \mu}=-\frac{\delta{\cal{L}}_{M}}{\delta A_{mn \mu}}+S^{\mu \nu [m}e^{n]}_{~\nu}.
\eqn 
On the other hand, a constraint equation will be achieved when the action is varied with respect to the tetrad
\bqn
\frac{\delta{\cal{L}}_M}{\delta e_{m \nu}}-D_{\mu}S^{\mu \nu m}=0.
\eqn
We would like to emphasize that $\delta e_{m \nu}$ is in fact $\eta^{\bar{m}\bar{n}}e_{\bar{n}\mu}\delta e_{\bar{m}i}$ in accordance with the arguments in section \ref{Sec:symmetry}.

One can now solve the latter constraint equation to find $S^{\mu \nu m}$ and substitute that in the former equation to find the whole Lorentz current. 
Inserting   
$
S^{\mu \nu i}= e_j^{~\mu}\xi^i\delta{\cal{L}}_M/\delta e_{j \nu},
$
into the constraint equation leads to a subsequent equation 
$
e_j^{~\mu}\frac{\delta{\cal{L}}_M}{\delta e_{j \nu}}\left(D_{\mu}\xi^i - e^i_{~\mu} \right)\simeq 0.
$
This equation implies that $D_{\mu}\xi^i = e^i_{~\mu} $ but only inside matter where $\delta{\cal{L}}_M/\delta e_{j \nu}$ is not zero. This equation should be valid for any matter distribution including a single particle. 
The solution for an observer sitting at ``the center of the particle'' is
\bqn
\lb{multiplier}
\xi^i(x)=
\lb{xiDef}
\begin{cases}
\eta^i_{~\alpha}(x^{\alpha}-X^{\alpha}) & x < \delta, \\
\xi^i_{_{+}} & x \geq \delta, 
\end{cases}
\eqn 
where $\delta$ is interpreted as the radius of the particle, $X$ refers to the center of the particle, and $\xi^i_{_{+}}$ is always multiplied by a zero. It only takes a transformation to find the solution in any other frame. 
In majority of classical systems 
$
\sum_{\text{particles}}\langle  \frac{\delta{\cal{L}}_{M}}{\delta A_{mn \mu}}\rangle = 0.
$
Moreover, since $\delta$ is of the ``size'' of particles, it can be integrated out of equation \eqref{fielddynamic} in macroscopic problems. It has been shown that the averaged source takes the following form \cite{Borzou2016_2}
\bqn
\lb{generalsource}
\tilde{J}_{\mu ij}=4\pi G \left( D_j T_{\mu i}-D_i T_{\mu j}\right),
\eqn
where 
\bqn
\lb{Eq:NewtonConstant}
G=\frac{g\delta ^2}{40 \pi},
\eqn
is Newton's gravitational constant. This is very similar to when the W boson's mass is integrated out of the electroweak theory and the Fermi constant $G_F=\frac{\sqrt{2}g_W^2}{8M_W^2}$ emerges. Both of the effective coupling constants are the dimensionless coupling constants of the underlying theory divided by the square of a large mass.  

The effective field equations can be written in terms of the space-time quantities if it is multiplied by two tetrads
\bqn
\nabla^{\nu}R_{\mu\nu\rho\sigma} = 4\pi G \left(\nabla_{\sigma}T_{\mu\rho}-\nabla_{\rho}T_{\mu\sigma}\right).
\eqn
This equation can be compared with general relativity if it is multiplied by $g^{\mu\rho}$
\bqn
\nabla^{\nu}R_{\nu\sigma} = 4\pi G \nabla_{\sigma}T,
\eqn
which, except a sign difference, is equivalent with the divergence of Einstein equation. The sign difference originates from the fact that the dynamical variable in LGT is proportional with a derivative of the dynamical variable in GR and maintains appropriate boundary conditions. 
This equivalence guarantees that LGT and GR have the same solutions to highly symmetric vacuums like the Schwarzschild and the de Sitter space-times \cite{Borzou2016}.


\section{Quantization of LGT in the Absence of Fermions}
\lb{Sec:Quantization_NoMatter}
In this section we would like to show that in the absence of matter the only difference between LGT and the Yang-Mills theories of the standard model of particle physics is their different gauge groups. 
Consequently, every theorem that is proved for general Yang-Mills theories is also valid for LGT in the vacuum, i.e. it is unitary and renormalizable to all orders.

In a given Yang-Mills theory the matter transforms according to some representation of a non-abelian Lie group 
$
\psi'(x) = e^{g t^a \alpha^a}\psi(x),
$
where $g$ is a constant, $a$ runs over the dimmension of the group, $\alpha^a$ are independent real parameters, and $t^a$ are generators of the Lie group.
From equation \eqref{Eq:LorTransform} it can be concluded that in LGT
\bqn
t^a=\left(S^{01},S^{02},S^{03},S^{12},S^{13},S^{23}\right),~ ~
\alpha^a=\left(\omega_{01},~\omega_{02},~\omega_{03},~\omega_{12},~\omega_{13},~\omega_{23}\right).
\eqn

In Yang-Mills theories the generators satisfy the Lie algebra
$
[t^a,t^b]=f^{abc}t^c,
$
where $f^{abc}$ are the structure constants. For LGT these are determined by equation \eqref{Eq:LorLieAlg}
\bqn
\lb{Eq:LGTStructureConstant}
f^{310}=f^{420}=-1, & ~ ~ , f^{031}=f^{521}=-1, & ~ ~, f^{042}=f^{152}=-1,\nb\\
f^{013}=f^{543}=-1, & ~ ~ , f^{024}=f^{354}=-1, & ~ ~, f^{125}=f^{435}=-1,
\eqn
which are the only independent non-zero ones. It should be noted that the structure constants are anti-symmetric in the first two indices. 
In a general Yang-Mills theory the covariant derivative is given by
\bqn
D_{\mu}\psi = \partial_{\mu} \psi -g A^a_{\mu}t^a\psi,
\eqn
where $A^a_{\mu}$ are the gauge fields and are determined for LGT through comparison with equation \eqref{Eq:LorCovDer}
\bqn
A^a_{\mu}=\left(A_{01\mu},~A_{02\mu},~A_{03\mu},~A_{12\mu},~A_{13\mu},~A_{23\mu}\right).
\eqn
In a given Yang-Mills theory the covariant derivative for a vector in the adjoint representation of the gauge group reads
\bqn
D_{\mu} \alpha^a = \partial_{\mu}\alpha^a - g f^{bca}A^b_{\mu}\alpha^c.
\eqn
One can easily confirm that this and equation \eqref{Eq:CovDerVec} are identical.
Finally the strength tensor in a general Yang-Mills theory is defined as 
\bqn
[D_{\mu},D_{\nu}]\psi = gF^a_{\mu\nu}t^a\psi.
\eqn
It is easy to show that 
\bqn
F^a_{\mu\nu} = \left(F_{\mu\nu 01},~F_{\mu\nu 02},~F_{\mu\nu 03},~F_{\mu\nu 12},~F_{\mu\nu 13},~F_{\mu\nu 23}\right).
\eqn
For a general Yang-Mills theory in the absence of matter the generating functional reads
\bqn
Z = \int D\bar{\eta}D\eta DA ~ e^{i \int e dx \left(-\frac{1}{4}F^a_{\mu\nu}F^{a\mu\nu}-\frac{1}{2\beta}\partial^{\mu}A^a_{\mu}\partial^{\nu}A^a_{\nu}-\bar{\eta}^a[\delta^{ab}\partial^2+\cev{\partial}^{\mu}f^{cba}A^c_{\mu}]\eta^b\right)},
\eqn
where $\eta^a$ are Grassmann fields. The fact that the metric, Christoffel symbols, and the determinant of the tetrad $e$ are background fields is a fundamental difference between a general Yang-Mills theory and general relativity. In LGT also these are background fields according to section \ref{Sec:symmetry}---only tetrad is not background in LGT but that only exists in the fermionic Lagrangian. 
It is now clear that this general generating functional includes LGT as well as it includes the theories in the standard model.
Therefore, the Feynman rules for LGT in vacuum can be readily read from the text books on the subject and are listed in appendix \ref{Ap:FeynRules}.

Since the dynamics of a Yang-Mills theory is determined by its generating functional, we have shown that LGT is just one of the possibilities of a general Yang-Mills theory. Therefore, any theorem that is proved for a general Yang-Mills theory is also valid for vacuum LGT. 
It has been shown that any Yang-Mills theory posseses a Slavnov-Taylor identity \cite{tHooft1971,Slavnov1972,Taylor1971}. 
The unitarity of the S matrix for arbitrary Yang-Mills theories have been proved \cite{tHooft1971}. 
Yang-Mills theories are also proved to be renormalizable to all orders \cite{veltman1972}.
Therefore, we can conclude that unlike general relativity that is not renormalizable and unitary at the same time even in the vacuum, LGT is simultaneously unitary and renormalizable to all orders of perturbation when matter is absent.

\section{Quantization in the Presence of Matter}
\lb{Sec:Quantization_WithMatter}
In section \ref{Sec:symmetry} it was shown that the metric, the Christoffel symbols, and the determinant of the metric---equivalently the determinant of the tetrad---are background fields under the dynamics of LGT. This was further used in section \ref{Sec:Quantization_NoMatter} to show that LGT in the absence of fermions is one of a general Yang-Mills theory that has been ruling the standard model of particle physics.

When fermions are present, there exists a difference between LGT and the Yang-Mills theories of the standard model. The tetrad---that is only present in the Lagrangian of fermions---is a background field in the latter but not in the former. 
In LGT this dependence is the result of the tetrad postulate which does not exist in any of the previously studied Yang-Mills theories. 
Therefore, in the presence of matter we can no longer use the proofs of unitarity and full renormalizability of arbitrary but unconstrained Yang-Mills theories. 
The proofs are not possible without first properly handling the extra constraint and deriving the Feynman rules. In this paper we focus on the latter and will postpone the former for future works.

The generating functional is
\bqn
Z = \int D\bar{\eta}D\eta DA D\bar{\psi}D\psi ~ e^{i \int e dx \left(-\frac{1}{4}F^a_{\mu\nu}F^{a\mu\nu}-\frac{1}{2\beta}\partial^{\mu}A^a_{\mu}\partial^{\nu}A^a_{\nu}-\bar{\eta}^a[\delta^{ab}\partial^2+\cev{\partial}^{\mu}f^{cba}A^c_{\mu}]\eta^b + \frac{i}{2}\bar{\psi}\gamma^{l}\tilde{e}_l^{~\mu}D_{\mu}\psi -
\frac{i}{2}\bar{\psi}\cev{D}_{\mu}\gamma^{l}\tilde{e}_l^{~\mu}\psi\right)},
\eqn
where the tilde on top of the tetrad is to remind us that it is not an independent field but a function of the gauge field. 

To proceed further, the following identity should be inserted in the generating functional above
\bqn
\int De~\delta\left(e - \tilde{e}\right) = \int DG\delta\left(G\right) = \int De~\text{det}\left(\frac{\delta G}{\delta e}\right)\delta\left(G\right) = \int D\lambda De~\text{det}\left(\frac{\delta G}{\delta e}\right)e^{-i\lambda^{i\nu}G_{i\nu}}=1,
\eqn
where $G_{i\nu} = D^{\mu}D_{\mu}e_{i\nu}=0$.
The determinant also can be sent into the exponential using Grassmann variables $c^{i\nu}$. The generating functional now reads
\bqn
Z &=& \int D\bar{\eta}D\eta DA D\bar{\psi}D\psi D\lambda De D\bar{c}Dc \times\nb\\
&&e^{i \int e dx \left(-\frac{1}{4}F^a_{\mu\nu}F^{a\mu\nu}-\frac{1}{2\beta}\partial^{\mu}A^a_{\mu}\partial^{\nu}A^a_{\nu}-\bar{\eta}^a[\delta^{ab}\partial^2+\cev{\partial}^{\mu}f^{cba}A^c_{\mu}]\eta^b + \frac{i}{2}\bar{\psi}\gamma^{l}e_l^{~\mu}D_{\mu}\psi -
\frac{i}{2}\bar{\psi}\cev{D}_{\mu}\gamma^{l}e_l^{~\mu}\psi - \lambda^{i\mu}D^2e_{i\mu} - \bar{c}^{i\nu}D^2 c_{i\nu}\right)}.
\eqn
This is the final generating functional for LGT and can be used to read the remaining Feynman rules. These are listed in appendix \ref{Ap:FeynRules_Matter}.

\section{Conclusions}
\lb{Sec:conclusion}
The Lagrangian of matter---fermions---is invariant under general covariance and local internal Lorentz transformations in the spinors. 
Energy-momentum tensor and the Lorentz current are the two associated conserved currents. 
Either of the two can be assumed to be the source of gravity. The latter leads to a spinless theory of gravity while the former to a spin 2 theory. 
Although there is yet no experimental measurement of the spin of gravitons, one can make a decision among the 
two spin options if gravity and the theories in the standard model of particle physics are assumed to follow the same principles. The particles in the latter are represented by the gauge fields of Yang-Mills theories. If this is going to hold in gravity, gravitons should be spinless. 

In this paper we have shown that in a spinless theory of gravity, the metric, its determinant, and the Christoffel symbols are all background fields. 
Moreover, in the absence of fermions the only difference between this theory and those in the standard model is that they are Yang-Mills theories of different gauge groups, i.e. they all belong to a general Yang-Mills theory. 
Consequently, every theorem of a general pure Yang-Mills theory is also valid for the spinless theory of gravity in vacuum. The most notable ones are the existence of a Slavnov-Taylor identity that holds to all orders, unitarity of the S matrix, and full renormalizability. 

In the presence of fermions there is a difference between the spinless theory of gravity and the theories in the standard model. The tetrad that is now present is a background field in the latter but not in the former. This is because in a theory of gravity the tetrad and the spin connection---the gauge field of the spinless theory---are dependent through the tetrad postulate. This means the spinless theory of gravity in the presence of matter is still a general Yang-Mills theory but with one extra constraint. 

A path integral quntization of the spinless theory of gravity---LGT---has been carried out as well. All the Feynman rules including those due to the tetrad postulate have been listed.

\appendix
\section{The Feynman Rules for a General Yang-Mills Theory in Vacuum}
\renewcommand{\theequation}{\Alph{section}.\arabic{equation}} \setcounter{equation}{0}
\lb{Ap:FeynRules}
In this section the Feynman rules for an arbitrary Yang-Mills theory is given. To find the rules in LGT one needs to substitute the structure constant in equation \eqref{Eq:LGTStructureConstant} into the following equations.

The propagator for the ghost field reads
\begin{fmffile}{GhostPropag}
\bqn
&&\nb\\
  \parbox{27mm}{\begin{fmfgraph*}(50,4)
    \fmfleft{l}
    \fmf{ghost,label=$k$}{l,r}
    \fmfright{r}
    \fmflabel{$b$}{l}
    \fmflabel{$a$}{r}
     \end{fmfgraph*}}
     & = & \frac{\delta^{ab}}{k^2}.
\eqn
\end{fmffile}
The interaction between the ghost field and the gauge field reads
\bqn
  \begin{fmffile}{GhostVertex}
      \setlength{\unitlength}{0.75cm}
      \parbox{30mm}{\begin{fmfgraph*}(3.5,3.5)
  \fmfright{i1,i2}
  \fmfleft{o}
  \fmf{ghost}{i1,v}
  \fmf{ghost,label=$p$}{v,o}
  \fmf{wiggly}{i2,v}
        \fmflabel{$c$}{i1}
        \fmflabel{$b$}{i2}
        \fmflabel{$a$}{o}
      \end{fmfgraph*}}
    \end{fmffile}
    = gf^{cba}p^{\mu},
\eqn
The propagator for the gauge field is
\begin{fmffile}{LGTPropag}
\bqn
  \parbox{27mm}{\begin{fmfgraph*}(50,4)
    \fmfleft{l}
    \fmf{wiggly,label=$k$}{l,r}
    \fmfright{r}
    \fmflabel{$b,\nu$}{l}
    \fmflabel{$a,\mu$}{r}
     \end{fmfgraph*}}
     & = & -\frac{\delta^{ab}}{k^2}\Big[g_{\mu\nu}-\left(1-\beta\right)\frac{k_{\mu}k_{\nu}}{k^2}\Big].
\eqn
\end{fmffile}
Finally the two interaction vertices are
\bqn
&&\nb\\
  &&\begin{fmffile}{A3Vertex}
      \setlength{\unitlength}{0.75cm}
      \parbox{30mm}{\begin{fmfgraph*}(3.5,3.5)
  \fmfright{i1,i2}
  \fmfleft{o}
  \fmf{wiggly,label=$k$,label.side=right}{i1,v}
  \fmf{wiggly,label=$p$}{o,v}
  \fmf{wiggly,label=$q$}{i2,v}
        \fmflabel{$c,\rho$}{i1}
        \fmflabel{$b,\nu$}{i2}
        \fmflabel{$a,\mu$}{o}
      \end{fmfgraph*}}
    \end{fmffile}
    =2gf^{cba}\Big[\left(k_{\mu}-q_{\mu}\right)g_{\nu\rho}+\left(p_{\nu}-k_{\nu}\right)g_{\mu\rho}+\left(q_{\rho}-p_{\rho}\right)g_{\mu\nu}\Big],
\eqn
and
\bqn
&&\nb\\
  &&\begin{fmffile}{A4Vertex}
      \setlength{\unitlength}{0.75cm}
      \parbox{30mm}{\begin{fmfgraph*}(3.5,3.5)
  \fmfright{i1,i2}
  \fmfleft{o1,o2}
  \fmf{wiggly}{i1,v}
  \fmf{wiggly}{o1,v}
  \fmf{wiggly}{o2,v}
  \fmf{wiggly}{i2,v}
        \fmflabel{$c,\rho$}{i1}
        \fmflabel{$b,\nu$}{i2}
        \fmflabel{$d,\sigma$}{o1}
        \fmflabel{$a,\mu$}{o2}
      \end{fmfgraph*}}
    \end{fmffile}
    =-ig^2\Big[f^{bea}f^{dec}\left(g_{\mu\rho}g_{\nu\sigma}-g_{\mu\sigma}g_{\nu\rho}\right)+f^{cea}f^{edb}\left(g_{\mu\sigma}g_{\rho\nu}-g_{\mu\nu}g_{\rho\sigma}\right)+f^{dea}f^{ceb}\left(g_{\mu\nu}g_{\sigma\rho}-g_{\mu\rho}g_{\sigma\nu}\right)\Big].\nb\\
    &&~
\eqn

\section{The Matter Dependent Feynman Rules of LGT}
\renewcommand{\theequation}{\Alph{section}.\arabic{equation}} \setcounter{equation}{0}
\lb{Ap:FeynRules_Matter}

Since the metric, its determinant and the Christoffel symbols are only background fields in LGT, and for the sake of simplicity we choose a flat space-time background, i.e. $<e_{i\mu}> = \eta_{i\mu}$ and consequently $e_{i\mu} = \eta_{i\mu} + h_{i\mu}$.

Fermion propagator reads
\begin{fmffile}{FermionPropag}
\bqn
  \parbox{27mm}{\begin{fmfgraph*}(50,4)
    \fmfleft{l}
    \fmf{fermion,label=$p$}{l,r}
    \fmfright{r}
     \end{fmfgraph*}}
     & = & \frac{p_{\mu}\gamma^{\mu}+m}{p^2-m^2}.
\eqn
\end{fmffile}
The vertex diagram for interaction of the gauge field and fermions is given by
\bqn
&&\nb\\
  &&\begin{fmffile}{Matter1A}
      \setlength{\unitlength}{0.75cm}
      \parbox{30mm}{\begin{fmfgraph*}(3.5,3.5)
  \fmfright{i1,i2}
  \fmfleft{o}
  \fmf{fermion}{i1,v}
  \fmf{wiggly}{o,v}
  \fmf{fermion}{v,i2}
        \fmflabel{$ab\lambda$}{o}
      \end{fmfgraph*}}
    \end{fmffile}
    =  \frac{g}{2}\eta_l^{~\lambda} \{\gamma^l,S^{ab}\} = - i \frac{g}{4} \varepsilon^{kij\mu}\gamma_k\gamma^5,
\eqn
where curly braces stand for anti-commutation divided by 2 and epsilon is the Levi-Civita symbol. 

Up to this point, every Feynman rule has been the same as those in an arbitrary Yang-Mills theory. 
The following Feynman diagrams are due to the tetrad postulate in LGT. The first is a propagator for the additional Grassmann variables $c^{i\nu}$
\begin{fmffile}{GhostPropag2}
\bqn
  \parbox{27mm}{\begin{fmfgraph*}(50,4)
    \fmfleft{l}
    \fmf{dbl_dots_arrow,label=$k$}{l,r}
    \fmfright{r}
    \fmflabel{$j\nu$}{l}
    \fmflabel{$i\mu$}{r}
     \end{fmfgraph*}}
     & = & \frac{\eta_{\mu\nu}\eta_{ij}}{k^2},
\eqn
\end{fmffile}
and a propagator for $e_{i\mu}$ 
\begin{fmffile}{e_Propag}
\bqn
  \parbox{27mm}{\begin{fmfgraph*}(50,4)
    \fmfleft{l}
    \fmf{dbl_zigzag,label=$k$}{l,r}
    \fmfright{r}
    \fmflabel{$j\nu$}{l}
    \fmflabel{$i\mu$}{r}
     \end{fmfgraph*}}
     & = & \frac{\eta_{\mu\nu}\eta_{ij}}{k^2},
\eqn
\end{fmffile}
and a propagarot for $\lambda_{i\mu}$
\begin{fmffile}{lambda_Propag}
\bqn
  \parbox{27mm}{\begin{fmfgraph*}(50,4)
    \fmfleft{l}
    \fmf{dashes,label=$k$}{l,r}
    \fmfright{r}
    \fmflabel{$j\nu$}{l}
    \fmflabel{$i\mu$}{r}
     \end{fmfgraph*}}
     & = & \frac{\eta_{\mu\nu}\eta_{ij}}{k^2}.
\eqn
\end{fmffile}
The remaining interaction diagrams are\\
\bqn
  \begin{fmffile}{Ghost2_Vertex}
      \setlength{\unitlength}{0.75cm}
      \parbox{30mm}{\begin{fmfgraph*}(3.5,3.5)
  \fmfright{i1,i2}
  \fmfleft{o}
  \fmf{dbl_dots_arrow,label=$q$,label.side=left}{i1,v}
  \fmf{dbl_dots_arrow,label=$p$}{v,o}
  \fmf{wiggly}{i2,v}
        \fmflabel{$n\sigma$}{i1}
        \fmflabel{$ab\lambda$}{i2}
        \fmflabel{$m\rho$}{o}
      \end{fmfgraph*}}
    \end{fmffile}
    = g\eta^{im}\delta^{ab}_{ij}\eta^{jn}\eta^{\sigma\rho}\left(p^{\lambda}+q^{\lambda}\right),
\eqn
where $\delta^{ab}_{ij}=\frac{1}{2}\left(\delta^a_i\delta^b_j-\delta^a_j\delta^b_i\right)$, and \\
\bqn
  \begin{fmffile}{Ghost2_Vertex2}
      \setlength{\unitlength}{0.75cm}
      \parbox{30mm}{\begin{fmfgraph*}(3.5,3.5)
  \fmfright{i1,i2}
  \fmfleft{o1,o2}
  \fmf{dbl_dots_arrow}{i1,v}
  \fmf{dbl_dots_arrow}{v,o1}
  \fmf{wiggly}{i2,v}
  \fmf{wiggly}{v,o2}
        \fmflabel{$n\sigma$}{i1}
        \fmflabel{$a_1b_1\lambda_1$}{i2}
        \fmflabel{$m\rho$}{o1}
        \fmflabel{$a_2b_2\lambda_2$}{o2}
      \end{fmfgraph*}}
    \end{fmffile}
    = -ig^2\eta^{im}\eta^{\sigma\rho}\eta^{kn}\eta^{jl}\eta^{\lambda_1\lambda_2}\left(\delta^{a_1b_1}_{il}\delta^{a_2b_2}_{jk}+
\delta^{a_1b_1}_{lk}\delta^{a_2b_2}_{ij}
    \right),
\eqn

\bqn
  \begin{fmffile}{Lambda_A}
      \setlength{\unitlength}{0.75cm}
      \parbox{30mm}{\begin{fmfgraph*}(3.5,3.5)
  \fmfright{i1}
  \fmfleft{o1}
  \fmf{dashes}{i1,v}
  \fmf{wiggly,label=$k$}{v,o1}
  \fmfdot{v}
        \fmflabel{$m\rho$}{i1}
        \fmflabel{$ab\lambda$}{o1}
      \end{fmfgraph*}}
    \end{fmffile}
    ~~~~~=g\eta^{im}\eta^{j\rho}\delta^{ab}_{ij}k^{\lambda},
\eqn

\bqn
  \begin{fmffile}{Lambda_A_A}
      \setlength{\unitlength}{0.75cm}
      \parbox{30mm}{\begin{fmfgraph*}(3.5,3.5)
  \fmfright{i2}
  \fmfleft{o1,o2}
  \fmf{dashes}{v,o1}
  \fmf{wiggly}{i2,v}
  \fmf{wiggly}{v,o2}
        \fmflabel{$a_1b_1\lambda_1$}{i2}
        \fmflabel{$m\rho$}{o1}
        \fmflabel{$a_2b_2\lambda_2$}{o2}
      \end{fmfgraph*}}
    \end{fmffile}
    ~~~~~~~~=-ig^2\eta^{im}\eta^{k\rho}\eta^{jl}\eta^{\lambda_1\lambda_2}\left(\delta^{a_1b_1}_{il}\delta^{a_2b_2}_{jk}+
\delta^{a_1b_1}_{lk}\delta^{a_2b_2}_{ij}
    \right),
\eqn

\bqn
  \begin{fmffile}{Lambda_Vertex}
      \setlength{\unitlength}{0.75cm}
      \parbox{30mm}{\begin{fmfgraph*}(3.5,3.5)
  \fmfright{i1,i2}
  \fmfleft{o}
  \fmf{dbl_zigzag,label=$q$,label.side=right}{i1,v}
  \fmf{dashes,label=$p$}{v,o}
  \fmf{wiggly}{i2,v}
        \fmflabel{$n\sigma$}{i1}
        \fmflabel{$ab\lambda$}{i2}
        \fmflabel{$m\rho$}{o}
      \end{fmfgraph*}}
    \end{fmffile}
    = g\eta^{im}\delta^{ab}_{ij}\eta^{jn}\eta^{\sigma\rho}\left(p^{\lambda}+q^{\lambda}\right),
\eqn

\bqn
  \begin{fmffile}{Lambda_Vertex2}
      \setlength{\unitlength}{0.75cm}
      \parbox{30mm}{\begin{fmfgraph*}(3.5,3.5)
  \fmfright{i1,i2}
  \fmfleft{o1,o2}
  \fmf{dbl_zigzag}{i1,v}
  \fmf{dashes}{v,o1}
  \fmf{wiggly}{i2,v}
  \fmf{wiggly}{v,o2}
        \fmflabel{$n\sigma$}{i1}
        \fmflabel{$a_1b_1\lambda_1$}{i2}
        \fmflabel{$m\rho$}{o1}
        \fmflabel{$a_2b_2\lambda_2$}{o2}
      \end{fmfgraph*}}
    \end{fmffile}
    = -ig^2\eta^{im}\eta^{\sigma\rho}\eta^{kn}\eta^{jl}\eta^{\lambda_1\lambda_2}\left(\delta^{a_1b_1}_{il}\delta^{a_2b_2}_{jk}+
\delta^{a_1b_1}_{lk}\delta^{a_2b_2}_{ij}
    \right),
\eqn
\bqn
&&\nb\\
  &&\begin{fmffile}{Matter_e}
      \setlength{\unitlength}{0.75cm}
      \parbox{30mm}{\begin{fmfgraph*}(3.5,3.5)
  \fmfright{i1,i2}
  \fmfleft{o}
  \fmf{fermion,label=$q$,label.side=right}{i1,v}
  \fmf{dbl_zigzag}{v,o}
  \fmf{fermion,label=$p$}{v,i2}
        \fmflabel{$m\nu$}{o}
      \end{fmfgraph*}}
    \end{fmffile}
    = \frac{i}{2}\gamma^m\left(p^{\nu}+q^{\nu}\right),
\eqn
 
\bqn
&&\nb\\
  &&\begin{fmffile}{Matter_e_A}
      \setlength{\unitlength}{0.75cm}
      \parbox{30mm}{\begin{fmfgraph*}(3.5,3.5)
  \fmfright{i1,i2}
  \fmfleft{o1,o2}
  \fmf{fermion}{i1,v}
  \fmf{dbl_zigzag}{v,o1}
  \fmf{fermion}{v,i2}
  \fmf{wiggly}{v,o2}
        \fmflabel{$ab\lambda$}{o2}
        \fmflabel{$m\nu$}{o1}
      \end{fmfgraph*}}
    \end{fmffile}
    = \frac{g}{2}\eta^{\lambda\nu}\{\gamma^m,S^{ab}\}.
\eqn

\end{document}